\documentclass[12pt]{article}
\usepackage{pic03}
\usepackage{hyperref}
\usepackage{url}
\usepackage{graphicx}

\begin{document}

%%%%%%%%%%%%%%%%%%  PSN: THBT01 %%%%%%%%%%%%%%%%%%%%%%%%%%%%%%%%%%%%
\title{\bf 
STATUS OF OBSERVATIONAL COSMOLOGY AND INFLATION
}
\author{
Laura Covi\\
{\em DESY Theory Group, Notkestrasse 85,}\\ 
{\em D-22603 Hamburg, Germany}
}
\maketitle

%
% photograph of author
%  This is where we will insert a photograph. To see what it would look like,
%  uncomment the following lines.
%
%\begin{figure}[h]
%\begin{center}
%
% include photograph for proceeding version
%
%\includegraphics
%[height=4.5cm]{einstein.eps}
%
% insert a fixed vertical spacing instead for the ArXiv preprint
%
\vspace{4.5cm}
%
%\end{center}
%\end{figure}

\baselineskip=14.5pt
\begin{abstract}
We review the latest developments in the determination of the
cosmological parameters from the measurement of the
Cosmic Microwave Background Radiation (CMBR) anisotropies
and of the Large Scale Structure (LSS) of the Universe.
We comment finally on the implications for the primordial 
spectrum and the consequences for inflationary models.
\end{abstract}
\newpage

\baselineskip=17pt

\section{Introduction}

It now more than apparent that we are in the 
era of precision cosmology:
during the last year we had an impressive progress,
the first detection of the polarization of the CMBR
by DASI and the precise determination of the 
CMBR anisotropies and confirmation of the DASI result 
by the satellite experiment WMAP.
Since the space at my disposal is limited, I will concentrate 
in these proceedings on the latest results and try 
to convey the status of observational cosmology after WMAP; 
I apologize in advance if I will appear to disregard all 
previous efforts, but they are covered for example
in the proceedings of the past editions of the PIC conference series.

This conference is mainly attended by particle physicists
not always familiar with the cosmological "jargon", so I will start
with a short review of Standard Cosmology and then follow up
with the latest observational highlights and the consequences
of such measurements for the cosmological parameters. 
The last section will be devoted instead to inflation and 
the attempt to relate the initial power spectrum with the one
predicted by a single-field inflationary model.

\section{Standard Cosmology}

The key assumption of our Standard Cosmological model
is the fact that the Universe is homogeneous and isotropic
on the very large scale. Thanks to this simplification 
we can write the background metric as a function of
very few parameters, in the Friedmann-Robertson-Walker
(FRW) form \cite{cosmo}:
\begin{equation}
ds^2 = dt^2 - R^2(t) \left[ {dr^2 \over 1- k r^2 } + r^2 d\Omega \right]
\end{equation}
where $R (t) $ is the {\it scale factor} of the universe and
the constant $k$ determines the spatial geometry 
($ k = +1, 0, -1$ for open, flat and closed universe respectively).

The evolution of the scale factor depends on the energy 
content and geometry of the universe and is given by the Friedmann 
equation,
\begin{equation}
H^2 = \left[ {\dot R (t) \over R(t) } \right]^2 = 
{8 \pi \over 3 M_{Pl}^2} \rho - {k\over R^2} + \Lambda.
\end{equation}
Here $M_{Pl} $ is the Planck mass, the energy density $\rho $ 
includes all the radiation and matter of the universe
(i.e. relativistic and non-relativistic particles), while
$\Lambda $ is the cosmological constant or "vacuum energy".  
$H (t)$ is the Hubble parameter and its present value 
$H_0 $ is called the Hubble constant; it is usually expressed 
via the adimensional quantity $h$,
which is $H_0$ in units of $ 100 $ km/s Mpc$^{-1}$.

In order to solve for the evolution of the scale factor,
it is also necessary to know the equation of state of the
different types of energy, defined as the ratio 
of pressure over density and use the first law of thermodynamics:
\begin{equation}
w = {P \over \rho} \quad\quad\quad\quad 
{d\rho\over dt} = - 3 H\; (\rho+P) =  - 3 H \rho\; (1+w).
\end{equation}
\begin{table}
\centering
\caption{ \it Equation of state, energy density as a function of $R$
and time dependence of $R$ and $H$ for different types of energy.
}
\[
\begin{array}[h]{| c | c c c c |}
      \hline
     \mbox{Type}  & w & \rho(R) & R(t) & H(t)   
      \\ \hline \hline
    \mbox{Generic} &  w & \propto R^{-3(1+w)} & \propto t^{2/(3(1+w))} & 
       2 (1+w)/ (3 t) 
    \\ \hline
    \mbox{Radiation} &  1/3 & \propto R^{-4} & \propto t^{1/2} & 1/(2 t) 
   \\ \hline
     \mbox{Matter} & 0 & \propto R^{-3}  & \propto t^{2/3} & 2/(3 t) 
   \\ \hline
       \Lambda & -1 & \mbox{const.}  & e^{H t} & \sqrt{\Lambda}
   \\ \hline
\end{array}
\]
\label{t-state}
\end{table}
In Table~\ref{t-state} are given the solutions for the energy density 
and the scale factor and Hubble parameter for different equations of
state. Note that the dependence of $H$ on time allows to distinguish 
between different types of energy.

Dividing by $H^2$, the Friedmann equation can be recast in the simple form
\begin{equation}
1 = \Omega_M + \Omega_k + \Omega_\Lambda
\end{equation}
where $\Omega_M = \rho/ \rho_c $, 
with $ \rho_c = {3 M_{Pl}\over 8\pi} H^2 $ being the critical density, 
while $\Omega_k = - k/(H^2  R^2)$ and 
$  \Omega_\Lambda = \Lambda/H^2 $.
Cosmologists therefore usually measure densities in
terms of the critical density as $\Omega_i h^2$, 
which is just the density in units of 
$\rho_c/h^2 = 1.879 \times 10^{-29} $~g/cm$^3$
$ = 1.054 \times 10^4$ eV/cm$^3$.
Also distances and time are measured by the redshift due to the 
cosmological expansion:
$1+z = \lambda_{obs}/\lambda_{em} = R_0/R(t_{em}) \geq 1 $,
where $\lambda_{em}, t_{em}$ are the wavelength and time
at emission and $\lambda_{obs}, R_0$ the observed wavelength
and the present scale factor respectively.

\subsection{Structure formation}

We have seen that the evolution of our universe is determined
by its energy content, but we have not yet explained how 
an isotropic and homogeneous state evolves into stars, galaxies,
cluster of galaxies. The Universe does not appear at all 
homogeneous on the small scale and this is due to the
gravitational attraction: once a small over-density appears, 
gravity causes it to grow and finally collapse into a 
bounded system.
So the basic ingredient for structure formation
is the presence of initial fluctuations in the 
density, that can in  later time act as seeds 
for the gravitational collapse \cite{infl-LSS}.

Assuming that the initial fluctuations were gaussian, 
their properties can be completely described in term
of their power spectrum 
\begin{equation}
{\cal P} (k) \simeq |\delta\tilde\rho(k)|^2 
\end{equation}
where $\delta \tilde\rho(k) $ is the Fourier transform
of the density contrast $\delta \rho(x) = \rho(x) - 
\bar\rho $. The dependence of the power spectrum from
the scale $k$ is very often parameterized by the power-law
$k^{n-1}$, with $n=1$ corresponding to the scale-invariant
case.

Since the fluctuations are very small (of the order of $10^{-5}$),
we can consider them as perturbations of the FRW metric and use 
perturbation theory to describe their dynamics.
Their evolution depends only on the cosmological parameters 
$h,\; \Omega_{tot},\; \Omega_M,\; \Omega_B$, the nature of Dark Matter
and the equation of state of the dominant component of the energy density.
Clearly the density fluctuations cannot grow as long as the pressure 
of the plasma counteracts the gravitational force and therefore
during radiation domination the system is still in the linear
regime and only oscillations in the plasma (the acoustic peaks !)
take place. Later, when matter dominates, the pressure drops 
to zero and the fluctuations can grow: structures start to form
and we enter the complicated non linear regime.

So any cosmological observation 
of the density contrast (at present or at recombination epoch)
contains in principle information on two class of quantities: 
the cosmological parameters describing the energy content of the 
universe and governing the evolution of the background metric
and the initial conditions for the fluctuations, also known as 
the primordial spectrum. It is not always easy to disentangle 
between the two and it is important to beware of
degeneracies and correlations between the different parameters.

\section{Cosmic Microwave Background Radiation}

The CMBR brings us information about the state of the universe
at the recombination epoch at about $z\simeq 1000$, when the
electrons were captured by the nuclei to form neutral atoms
and radiation decoupled. 
The photons that reach us now had their last
scattering at that time. Density fluctuation in the plasma in 
thermal equilibrium gave rise to temperature fluctuations, since the
denser regions were hotter. So the temperature anisotropies
in the CMBR bring us direct evidence of the density contrast
at recombination.

It is traditional to express the temperature anisotropies into
spherical harmonics functions \cite{CMBR}:
\begin{equation}
\Delta T (\theta,\phi) = \sum_{\ell, m} a_{\ell m} 
Y_{\ell m} (\theta,\phi)
\end{equation}
and obtain the temperature anisotropies as a function of the
multi-pole number 
\begin{equation}
C_\ell \equiv {1\over 2\ell +1} \sum_m |a_{\ell m}|^2.
\end{equation}

Note that the multi-pole number $\ell $ corresponds to a
particular length $ \lambda = k^{-1} \propto \ell^{-1} $ on the
last scattering surface. Since at this epoch we are
still in the linear regime, the equation of motion for the 
single scales are independent and can be solved, e.g.
using numerical codes like CMBFAST \cite{cmbfast}.  
In very broad terms three different
behaviours are present, depending on the wavelength: 
large scales (i.e. small multi-poles) do not oscillate, 
but feel the presence of the gravitational potential; 
scales equal or smaller than the 
sound horizon oscillate and the first acoustic peak 
corresponds to the scale of the horizon who has just 
completed one compression before recombination, 
while even peaks are instead rarefaction
peaks; finally the very small scale (large multi-poles)
oscillations are damped. For a detailed description 
of the physics of the CMBR see \cite{CMBR,hu-web}.

\subsection{Polarization}

Another interesting property of the CMBR, which has
been firstly measured in the last year, is that it
is partially linearly polarized. Such polarization is due
to the fact that Thomson scattering tends to produce
preferentially a final photon polarized in the same 
direction as the initial photon \cite{polar,hu-web}. 
It is clear then 
that if the plasma would be completely isotropic,
no net polarization would arise. But due to the
density fluctuations, the local velocity field
of the photons (i.e. the flux) in the rest system 
of the electrons is not homogeneous. Then the polarization 
reflects the local quadrupole in the
velocity field, which is anti-correlated with the
temperature anisotropies (the velocity is zero
at maximal compression or rarefaction). 
The polarization can in general be decomposed into
two different modes, the curl-free mode $E$ and the
curl mode $B$. Since the $B$ mode is a pseudo-vector
quantity, its cross correlations with $T$ and $E$
(which are a scalar and a vector) vanish. 
The cross correlation between the temperature 
anisotropies and the $E$ mode of polarization is 
instead non-vanishing.
We can therefore describe completely the CMBR temperature
and polarization anisotropies via four independent correlations,
the $<TT> $ correlation which corresponds to $ C_l$
introduced in the previous subsection, the mixed
correlation between temperature and
$E-$mode polarization anisotropies $<TE> $, and finally the pure
$E-$ and $B-$mode correlations  $<EE> $ and $<BB>$.

It is important to say that scalar density fluctuations
excite predominantly $E$-mode polarization, while
tensor fluctuations (as expected from gravity
waves) excite the $E$- and $B$-modes at the same level.
Once both polarization modes will be detected, it will
be clear how large is the contribution of 
the tensor fluctuations.

\subsection{DASI}

DASI (Degree Angular Scale Interferometer) is an interferometer
experiment located at the South Pole, who first detected
the CMBR polarization. The measurement was announced last
year during the COSMO-02 conference (you can see the announcement
live on the web-site \cite{DASI-video}).
For the detailed results and plots, I refer to the DASI web-site 
\cite{DASI-web} and their publications \cite{DASI-papers}; 
the experiment measured all the four correlation
in the multi-pole range between $200$ and $800$ and obtained 
evidence for a non-zero $E$-mode at 4.9$\sigma$.
Their signal is in agreement with the expected $E$-mode
polarization produced by scalar density fluctuations and
with the measured temperature anisotropies.
The $B-$mode instead is consistent with zero.

\subsection{WMAP}

WMAP (Wilkinson Microwave Anisotropy Probe) is a satellite
experiment launched by NASA in 2001 (for the details of the
mission, please look at their very exhaustive web-page
\cite{WMAP-web}). The satellite completed the first full sky 
observation in April '02 and the first data release based on that
sky map took place this year, in February. The data and pictures 
are publicly available on the web-portal LAMBDA \cite{WMAP-lambda}.
WMAP is continuing to take data and so there is more to come
in the future. 

The WMAP team is measuring the intensity of radiation in 5 different
bands and then the five maps (subtracting the dipole component and 
the Milky Way) are combined to obtain the full
sky map of the temperature anisotropies. 
WMAP also measures the cross correlation between
temperature and $E$-mode polarization anisotropies \cite{WMAP-kogut}.

The WMAP data can reach the multi-pole $\ell\simeq 900$, 
up to the third acoustic peak; to extend to the higher 
multi-poles, to $\ell \simeq 1700$, the WMAP team included in their 
analysis the data of other two CMBR ground-based experiments,
ACBAR (Arcminute Cosmology Bolometer Array Receiver) \cite{acbar} 
and CBI (Cosmic Background Imager) \cite{cbi}.

\section{Large Scale Structure (LSS)}

Information on the density contrast can also be obtained from the
distribution of galaxies in our universe. The main assumption in 
this case is that the visible matter follows the distribution of 
the invisible Dark Matter. The unknown difference between the
two distribution is usually parameterized by the bias parameter.
Also it is necessary to correct for the non-linearity in the
evolution of the small scale perturbations to extract
the present linear spectrum, which 
allows to access directly the primordial one.

Present surveys include the 2 degree Field Galaxy Redshift Survey 
(2dF GRS) just completed, which released recently data 
about 270.000 galaxies  \cite{2dF-web}.
An even larger survey is ongoing, the Sloan Digital Sky Survey 
(SDSS), which aims at 1 million of galaxies in one quarter of the sky
\cite{SDSS-web}.
From the distribution of the galaxies in the sky one can
obtain the two point correlation function and the density
contrast power spectrum.

Other ways to measure the density contrast rely on 
using photons of distant objects as a probe of the 
intervening matter or gas densities. 
Lyman $\alpha $ forest data measure the absorption lines
in the spectra of distant quasars caused by intergalactic
hydrogen and estimate the cosmic gas distribution out to
large distances \cite{Ly-alpha}. 
This method makes possible to access also the
power spectrum at the very small scales, but its systematics
are still under debate, since the power spectrum estimation 
relies on modeling and must be 
corrected for non-linearities \cite{Ly-alpha-seljak}.
Yet another way of accessing the matter distribution is
weak gravitational lensing, which
measures the shear (distortion) in the images of distant
objects due to the gravitational potential of the intervening
matter \cite{weak-lens}. Weak lensing is sensitive to the 
total matter distribution along the line of sight, without any bias.
Other methods to obtain informations on the matter distribution
and the power spectrum are X-rays measurements \cite{X-ray} and 
peculiar velocities \cite{cosmo}.

One recent development about measurements of the matter
power spectrum is the fact that the results from CMBR determinations
and from the different LSS methods are now overlapping with each
other and cover continously all the scales between
the horizon size, about $10^4\; h^{-1} $ Mpc, and $1\; h^{-1} $ Mpc 
(see e.g. \cite{tegmark} for a compilation of data on the
power spectrum just before WMAP).

\section{The cosmological parameters from WMAP and LSS}

\subsection{Total energy density and matter density}

As discussed previously, the position of the first peak 
in the CMBR power spectrum corresponds to a wavelength equal 
to the sound horizon at the surface of last scattering.
Since the sound horizon for a relativistic plasma is
known, the angular position of the first peak measures 
directly the geometry of the universe, i.e. $\Omega_k $.
It is usual to express such a measure in terms of the
the total energy density $\Omega_{tot} = 1 - \Omega_k $.
From a global fit including LSS data and a prior on the
value of $H_0$, WMAP obtains 
$\Omega_{tot} = 1.02 \pm 0.02 $\footnote{All errors are 
$1\sigma $ if it is not explicitely stated otherwise.}
\cite{WMAP-bennet},
completely in agreement with the previous result of
Boomerang \cite{boom} and MAXIMA \cite{maxima}.

The dependence of the temperature anisotropies on the 
matter density is more involved and the quantity determined
from the global fit is $\Omega_M h^2 = 0.135^{+0.008}_{-0.009} $
\cite{WMAP-spergel}. Using the best fit value for the 
Hubble constant $h = 0.71^{+0.04}_{-0.03}$, this gives
$\Omega_M = 0.27 \pm 0.04$ \cite{WMAP-spergel}. 
Together with the measurement of the total energy density,
we obtain then $\Omega_\Lambda \simeq 0.7 $, as
measured independently from Supernova IA (SN IA) data 
\cite{SN-IA-web, SN-IA-papers}.

\subsection{Baryon density}

The baryon matter present in the plasma at recombination
changes the dynamics of the oscillations, enhancing 
the compressions and suppressing the rarefactions.
Therefore a very precise measurement of the baryon density 
comes from the comparison between odd and even CMBR peak heights.
The WMAP collaboration obtains 
$ \Omega_b h^2 = 0.0224 \pm 0.0009 $ \cite{WMAP-spergel}.
This values is much more precise and fully consistent with 
the one computed from the observed Deuterium abundance with standard
Big Bang Nucleosynthesis (BBN). 
Still the $^4He$ and $^7Li$ abundances are not so concordant,
as discussed in detail in \cite{BBN}. It is still open if
such discrepancies are due to systematics, non standard BBN or else.

\subsection{Hot Dark Matter and neutrinos}

Light particles with mass of the order of eV, 
which remain relativistic down to late times, constitute what
is called Hot Dark Matter (HDM). Their main characteristic 
for what regards structure formation is the fact that they have
a relatively large free-streaming length: for a massive
non-interacting particle the free-streaming length is
given by $ \lambda_{FS} \simeq 1200/m_{\mbox{\small eV}}\;\; \mbox{Mpc} $,
where $m_{\mbox{\small eV}} $ is the mass of the particle
in eV.  The presence of such free streaming suppresses 
the formation of bound systems of sizes smaller than $\lambda_{FS}$,
i.e. it suppresses the power spectrum of the small scales. 
This effect is absent for Cold Dark Matter, which has 
practically zero free-streaming length.
For this reason the HDM density can be estimated
from the power spectrum at small scales, which is measured 
by LSS data, in particular Lyman $\alpha $ data.
The CMBR measurements are less sensitive to HDM since they 
sample larger scales than $\lambda_{FS}$, but they are useful
for setting the other parameters in the game.

Combining CMBR, 2dF GRS and Lyman $\alpha $ data,
 the WMAP collaboration obtains  at 98\% CL
the bound  $ \Omega_\nu h^2 \leq  0.0076 $ \cite{WMAP-spergel},
which for 3 degenerate neutrinos with thermal distribution can 
be translated into
$ m_\nu \leq  0.23\, \mbox{eV}  $ at 98\% CL.
Note that the limit strongly depends on the cosmological parameters,
the number of sterile neutrinos and their thermalization, 
as discussed by \cite{hannestad}.
There a more conservative, but qualitatively not much different
bound is obtained as:
$ \sum m_\nu \leq 1.01\; (1.38) \;\mbox{eV} \;\;\; \mbox{for} \;\;\;
N_\nu = 3\; (4) $.
Not surprisingly, the bound is driven mostly by the Lyman $\alpha $
data, which sample the smaller scales; substituting those
with X-rays data, \cite{allen} finds instead evidence
for a non zero neutrino mass, 
$ \sum m_\nu = 0.56^{+0.30}_{-0.26} \;\mbox{eV} $.
Further analysis and data are needed to clarify the issue.

Note anyway that the effect by HDM of suppressing the power 
spectrum at small scales could be partially mimicked by the "running" 
of the spectral index of the primordial spectrum or by the presence
of quintessence at early times~\cite{early-Q}.

\subsection{Equation of state of the Dark Energy} 

The WMAP team also performed a global fit for the equation of state
of the Dark Energy combining CMBR data with the 2dF GRS and SN IA 
data and obtained the bound $w \leq -0.78 $ at 95\% CL 
\cite{WMAP-spergel} using a constant $w \leq 1$. 

\subsection{Reionization}

A surprising result of the WMAP polarization data is the
high signal at low multi-poles $\ell < 10 $. 
Such polarization on very large scales cannot be generated at 
recombination, but is instead due to Thomson scattering at a
much later epoch, when the universe was re-ionized. 
The epoch of reionization took place after the gravitational 
collapse, when the light of the first stars ionized 
the hydrogen clouds.
The effect of the presence of ionized gas on the temperature
anisotropies is to reduce the spectrum by a factor $e^{-2 \tau}$, 
where $\tau $ is the optical depth for a photon traveling in 
the ionized medium from the epoch of reionization to today.
Until the WMAP result, the optical depth $\tau $ was thought
to be very near to zero, consistent with the
models which describe the formation of the early stars
at $z \simeq 6 $. The polarization data instead give
$\tau = 0.17 \pm 0.06 $, hinting at reionization
happening at much earlier times, $z \simeq 20$ \cite{WMAP-kogut}.
This independent determination of reionization via the
polarization is very important because it breaks the 
degeneracy between the spectral index $n$ and $\tau$ present 
in the temperature anisotropies.

\section{Inflationary predictions}

Inflation is a period of exponential expansion, which is 
usually introduced before the Standard Cosmology and
solves some of its problems about initial conditions \cite{infl}.
The exponential expansion is driven by the effective cosmological
constant due to the displacement of a scalar field from the 
minimum of its potential.
In order for the potential energy to give this effect,
the scalar field must have negligible kinetic energy and
descend very slowly towards the minimum, 
i.e. the inflaton potential $V$
has to satisfy the slow roll conditions:
\begin{equation}
\epsilon = {M_{Pl}^2 \over 2} \left( {V'\over V }\right)^2 \ll 1
\quad\quad | \eta | = M_{Pl}^2 \left| {V''\over V} \right| \ll 1,
\end{equation} 
where the primes denote first and second derivative w.r.t. 
the inflaton field.
Inflation ends when the scalar field starts to roll faster
and finally to oscillate around the minimum; then it decays
producing radiation and reheating the universe.

Many different models of inflation have been proposed and
studied \cite{lr98}, all involving physics beyond the
Standard Model, and one important question is to confront
them with the data and try to gain insight on the new physics.

\subsection{Comparison with the data}

The CMBR observations agree very well with the general
inflationary predictions for single field models \cite{infl-LSS},
as we will discuss in detail.

\begin{itemize}
\item{Inflation produces in general a spatially flat universe, 
i.e. $\Omega_{tot} = 1 $, in perfect agreement with the WMAP
measure of $\Omega_{tot} = 1.02\pm 0.02 $ \cite{WMAP-spergel}.
}
\item{The primordial fluctuations are generated by the quantum
fluctuation of the inflaton field, which is practically massless
and non-interacting, and therefore they are gaussian and adiabatic;
the measured power spectrum is consistent with gaussianity 
\cite{WMAP-komatsu} and  adiabaticity
(the fraction of isocurvature perturbations $f_{iso} \leq 0.33$
at 95\% CL)
\cite{WMAP-peiris}.
}
\item{The slow roll approximation predicts a nearly scale-invariant 
spectrum of the scalar primordial perturbations
related to the inflaton potential as following:
\begin{equation}
{\cal P} (k) = \left. {1\over 12 \pi^2 M_{Pl}^6 }{V^3\over (V')^2}
\right|_{k=H R} 
\end{equation}
where the l.h.s. is evaluated at the inflaton value
corresponding to the time when the physical scale $R/k$ was 
equal to the horizon $H^{-1}$. Due to the slow roll of the
inflaton field, the dependence from $k$ is expected to be weak. 
In fact the spectral index is at lowest order in slow roll:
\begin{equation}
n(k) - 1 = \left. {d\log({\cal P_R})\over d\log(k)} \right|_{k=H R}
= 2 \eta - 6 \epsilon\; ,
\end{equation}
so inflation predicts a very small deviation from the 
Harrison-Zeldovich scale invariant case $n=1$.
The WMAP data are actually consistent with a spectral index 
equal to one, $ n = 0.99\pm 0.4 $ in \cite{WMAP-spergel}. 
A similar result comes also from other data analysis \cite{n-b1,n-b2,n-k,n-ll}.
}
\item{A surprising hint from WMAP is the preference of the
data for a "running" of the spectral index $n$, which means a
non trivial $k$ dependence. Expanding $n$ in a Taylor series
around a reference scale $k_0$,
\begin{equation}
n(k) = n(k_0) + n'(k_0) \ln \left( {k\over k_0} \right) + ...
\end{equation} 
and neglecting higher order terms, the WMAP team obtains 
the best fit value, $n'(k_0) = -0.055^{+0.028}_{-0.029} $
for $k_0 = 0.002 \mbox{Mpc}^{-1}$ \cite{WMAP-peiris}.

The central value for $n'$ appears too large to be 
consistent with the single field inflationary prediction,
which is
\begin{equation} 
n'(k) = -{2\over 3} \left[ (n-1)^2 - 4 \eta^2 \right] - 2 \xi^2,
\end{equation}
where $\xi^2 $ is a second order slow roll parameter
$ \xi^2 = M_{Pl}^4 {V' V'''\over V^2}$.
In fact the only way to accommodate large running for
small $(n-1)^2 \simeq 4 \eta^2 \leq 0.02 $ is to assume a 
strange potential with unnaturally large $\xi^2$ \cite{run}. 
Of course the value for $n'$ is consistent with zero at 
the $2\sigma$ level \cite{WMAP-peiris}, or even less 
\cite{n-b1,n-b2,n-k,n-ll}, so further investigations and statistics 
are required to settle the question. 
}

\item{Tensor perturbations (primordial gravity waves) have 
instead power spectrum:
\begin{equation}
 {\cal P}_{grav} (k) =  \left. {1\over 6 \pi^2} 
{V\over M_{Pl}^4} \right|_{k=H R}
\end{equation}
and spectral index
\begin{equation}
n_{grav}(k) =  
\left. {d\log({\cal P}_{grav})\over d\log(k)} \right|_{k=H R} = 
- 2 \epsilon .
\end{equation}
So if the scale of inflation potential is much lower than the Planck
mass, the contribution of tensor perturbations can be negligible.
At the moment the data show no evidence for tensor perturbation, 
and for $r ={\cal P}_{grav}/{\cal P_R} $ one finds 
$ r < 0.90 $ at 95\% CL \cite{WMAP-spergel}.
}
\end{itemize}

\section{Conclusions and outlook}

The era of precision cosmology continues and in this year
we have seen the "concordance" $\Lambda$CDM universe
confirmed by the new WMAP data. The cosmological parameters 
are nowadays measured to the level of few per cent, unthinkable
precision up to a couple of years ago.
This and the overlap between CMBR and LSS data for the power spectrum 
allow to put better constraints on the models of structure 
formation, as demonstrated by the strong bound on the energy 
density in neutrinos.

The polarization of the CMBR has been for the
first time detected in the past year and this result
confirms beautifully our understanding of the physics
of acoustic peaks and gives an independent measure of the
reionization.

For what regard the primordial fluctuations,
the simple single field inflationary paradigm with negligible
tensor perturbation is at the moment sufficient to describe 
the data. 

Some small discrepancies in the concordance picture 
are appearing, e.g. BBN, the running of the spectral index,
reionization, etc..., and are the object of present studies.
We are looking forward to the next year with more data 
to come !

\section{Acknowledgments}

I would like to thank the organizers for the invitation
to speak and the exciting atmosphere at the Workshop.

The bibliography here includes basically reviews and papers 
published in the last year in order to highlight the recent 
developments, it is surely not exhaustive nor complete. 
I apologize for the absences and I point the interested 
readers to the references of the references, in particular
\cite{cosmo,infl-LSS,CMBR,polar,lr98}.
In the case of on-going experiments, I decided to give the reference
not only of the publications, but also of the collaboration web-sites, 
where frequently updated informations and the latest achievements 
can be found.

\end{document}